\magnification=\magstep1
\tolerance=500
\bigskip
\rightline{23 July, 2017}
\bigskip
\centerline{\bf Canonical Transformation of Potential Model Hamiltonian} \centerline{\bf Mechanics to Geometrical Form I}
\bigskip
\centerline{ Y.Strauss$^{1,2}$, L.P. Horwitz$^{2,3,4}$, A. Yahalom$^2$, J. Levitan$^2$}
\bigskip
\centerline{${}^1$ Ben Gurion University, Beer Sheva, 84105 Israel}
\centerline{${}^2$ Ariel University, Ariel, 40700 Israel}
\centerline{${}^3$ Tel Aviv University, Ramat Aviv, 69978 Israel}
\centerline{${}^4$ Bar Ilan University, Ramat Gan, 52900 Israel}

\bigskip
\noindent{\it Abstract}
\par Using the methods of symplectic geometry, we establish the existence of a canonical transformation from potential model Hamiltonians of standard  form in a Euclidean space to an equivalent geometrical form on a manifold, where the corresponding motions are along geodesic curves. The advantage of this representation is that it admits the computation of geodesic deviation as a test for local stability, shown in recent previous studies to be a very effective criterion for the stability of the orbits generated by the potential model Hamiltonian. We describe here an algorithm for finding the generating function for the canonical transformation and describe some of the properties of this mapping under local diffeomorphisms.   We give a convergence proof for this algorithm for the one dimensional case, and provide a precise geometric formulation of geodesic deviation which relates the stability of the motion in the geometric form to that of  the Hamiltonian standard form.  We discuss the relation of bounded domains in the two representations for which  Morse theory would be applicable.  Numerical computations for some interesting examples will be presented in succeeding papers. 
\bigskip
\noindent{\bf  1. Introduction}
\par This paper is concerned with the development of a new method for embedding the motion generated by a classical Hamiltonian of standard form into a Hamiltonian defined  by a bilinear form on momenta with coordinate dependent coefficients (forming an invertible matrix) by means of a canonical transformation. This type of Hamiltonian, which we shall call {\it geometric}, results, by applying Hamilton's equations, in equations of motion of geodesic form. The coefficients of the resulting bilinear form in velocities can be considered to be a connection form associated with the coefficients in the momenta in the geometric Hamiltonian considered as a {\it metric} on the corresponding coordinates. The advantage of this result, which may be considered to be an embedding of the motion induced by the original Hamiltonian into an auxiliary space for which the motion is governed by a geodesic structure, is that the deviation of geodesics on such a manifold (involving higher order derivatives than the usual Lyapunov criteria) can provide a very sensitive test of the stability of the original Hamiltonian motion.
\par In previous work, an {\it ad hoc} construction of a geometrical
embedding using a conformal metric [1]was introduced.   Casetti and Pettini  [2] have investigated the application of the Jacobi metric and the extension of the analysis of the resulting Jacobi equations along a geodesic curve in terms of a parametric oscillator; such a procedure could be applied to the constuction we discuss here as well.  The relation of the stability of geometric motions generated by metric models previously considered to those of the motion generated by the original Hamiltonian is generally, however,  difficult to establish. The transformation that we shall construct here preserves a strong relation with the original motion due to its canonical structure.  
\par The methods we shall use are fundamentally geometric, involving the properties of symplectic manifolds which enable the definition and construction of the canonical transformation without using the standard Lagrangian methods. These geometric methods provide a rigorous framework for this construction, which makes accessible a more complete understanding of the dynamics.
\par  The theory of the stability of Hamiltonian dynamical systems has been discussed in depth, for example, in the books of Ar'nold [3],Guckenheimer and Holmes [4], and recently by DiBenedetto [5]. In his discussion of stability, Gutzwiller [6](see also Miller and Curtiss [7]) discusses the example of a Hamiltonian of geometric type, where the Hamiltonian, instead of the standard expression
$$ H(q,p)= { p^2 \over 2m} + V(q), \eqno(1.1))$$
 has the form (in two or more dimensions),
$$H_G(x,\pi)= {1 \over 2}g_{ij}(x)\pi^i \pi^j, \eqno(1.2)$$
with indices summed.\footnote{*}{We use the convention, differing from that of the standard literature on differential geometry, of denoting coordinates with lower indices and momenta with upper indices, to conform with the usage in [1].}    In  one dimension, $g(x)$ would be just a scalar
function, but, as we shall see, is still of interest. We shall call
such a structure {\it geometrical}. We shall call the space of the
standard variables $\{q,p\}$ the {\it Hamilton space.}
The application of Hamilton's equations to Eq.$(1.2)$ results in a geodesic type equation
$$ {\ddot x}_\ell = - {\Gamma_\ell}^{mn}{\dot x}_m {\dot x}_n,
\eqno(1.3)$$
where the coefficients have the structure of a connection form
(here, $g^{ij}$ is the inverse of $g_{ij}$)
$$ \Gamma_\ell^{mn} = {1\over 2} g_{\ell k} \bigl\{ {\partial g^{km} \over
\partial x_n}+  {\partial g^{k n} \over
\partial x_m}- {\partial g^{n m} \over
\partial x_k} \bigr\}. \eqno(1.4)$$
This connection form is {\it compatible} with the metric $g_{ij}(x)$
by construction, {\it i.e.}, the covariant derivative of of $g_{ij}$ constructed with
the $\Gamma_\ell^{mn}$ of Eq. $(1.3)$  vanishes, and we recognize that the dynamics
generated on the coordinates $\{x\}$ is a geodesic flow.  It can carry,
moreover, a tensor structure which may be inferred from the
requirement of invariance of the form $(1.2)$ under local coordinate transformations.
\par The stability of such a system may be tested by studying the {\it geodesic deviation}, {\it i.e.}, by studying what happens when one shifts to a nearby geodesic curve, corresponding to a local change in initial conditions.  The resulting separation of the two geodesic curves provides a very sensitive test of stability (see Gutzwiller [6], and for its application to general relativity, Weinberg [8]). An exponentially growing  deviation is characteristic of local instability, and may lead to chaotic behavior of the global motion.
\par In order to obtain a criterion in the case of a standard
Hamiltonian of the form $(1.1)$, Horwitz {\it et al} [1] constructed
an {\it ad hoc} transformation of this Hamiltonian to a Hamiltonian of the form $(1.2)$ by defining the metric as
$$   g_{\ell k}(x)= \delta_{\ell k}\phi(x),\eqno(1.5)$$
where (with a relation between $x$ and $q$ to be explained below)
$$\phi(x) = {E \over E-V(q)} \equiv F(q)\eqno(1.6),$$ 
and $E$ is taken to be the assumed common (conserved) value of $H$ and
$H_G$.
\par The motion induced on the coordinates $\{x\}$ by $H_G$ , after the local tangent space transformation
 ${\dot y}^k =g^{k\ell}(x) {\dot x}_\ell$, results in a {\it geometric embedding} of the original Hamiltonian motion for which the geodesic deviation gives a sensitive diagnostic criterion for the stability of the original Hamiltonian motion [1, 9, 10].  The condition of dynamical equivalence of the two systems,
based on enforcing equal values of the momenta at all times (the
transformation is not necessarily canonical), provides a constraint that establishes a correspondence between the coordinatizations $\{x\}$ and $\{q\}$ in the sense that $\phi(x)$ can be expressed as a series expansion in $F(q)$ and its derivatives, and conversely, $F(q)$ can be expressed as a series expansion in $\phi(x)$ and its derivatives, in a common domain of analyticity [9]; in this way, all derivatives of $\phi(x)$ can be expressed in terms of derivatives of $F(q)$, and conversely.
\par The remarkable success of this method has not yet been explained,
although some insights were provided in [12]. In the theory of
symplectic manifolds [13], a well defined mechanism exists for transforming a Hamiltonian of the form $(1.1)$ to that of $(1.2)$ (with a possibly conformal metric) by a rigorous canonical transformation, admitting the use of geodesic deviation to determine stability, which would then be clearly associated with the original Hamiltonian motion. We shall define this theory, and describe some of its properties, in this paper.
\par  We remark that in an analysis [14] of the geodesic deviation treated as a parametric oscillator, a procedure of second quantization was carried out providing an interpretation of excitation modes for the instability in a ``medium'' represented by the background Hamiltonian motion. This interpretation would be applicable to the results of the construction we present here as well.  
\par In the following, we describe this mapping and an algorithm for obtaining solutions.
We give a convergence proof for the recurrence relations for the generating function in the one dimensional case which appears
 to be applicable to the general {\it n}-dimensional case. Although the algorithm for the construction is clearly effective (and convergent), its realization requires considerable computation for specific applications, which we shall carry out in succeeding publications. The resulting programs could then be applied to a wide class of systems to provide stability criteria without exhaustive simulation; the local criteria to be developed could, furthermore, be used for the control of intrinsically chaotic systems [10].
 \par In this paper we discuss some general properties of the
 framework. In Section 2, we give the basic mathematical
 methods in terms of the geometry of symplectic manifolds.
 \par A central motivation for our construction is to make available
  the  study of  stability by means of geodesic
 deviation. This procedure is studied in Section 3, in terms of geometric methods, making clear
 the relation between stability in the geometric manifold and the
 original Hamiltonian motion.    
 \par In Section 4, an algorithm is described for solving 
 the  nonlinear equations for the generating function of the canonical
 transformation.
 In Section 5, we study this algorithm for the one
 dimensional case, and prove convergence of the series expansions, under certain assumptions in
 Section 6. The series expansions that we obtain can be studied by methods of Fourier series representations; the nonlinearity leads to convolutions of analytic functions (see,for example Hille [15]) that may offer approximation methods that could be useful in studying specific cases. We plan to discuss this topic in a future publication. 
 \par Since the iterative 
expansions for the generating function could be expected to
 have only bounded domains of convergence, we consider, in Section 7, the possibility of
 shifting the origin of the expansion in general dimension,  As for the analytic continuation of a function of a complex variable, this procedure can extend the definition of the generating function to a maximal domain.  
\par Since the image space of the symplectomorphism has geometrical
 structure, it is natural to study its properties under local
 diffeomorphisms. A local change of variable alters the structure
 of the symplectomorphism. We study the effect of such diffeomorphims on the generating function (holding the original Euclidean
 variables fixed) in Section 8.
\par   Further mathematical implications, such as relations to Morse
 theory ({\it e.g.} [16],[17]), are briefly discussed in Section 9; a more extended development of this topic will be given in a succeeding publication.   
\bigskip
\noindent{\bf 2. Basic mathematical formulation}
\bigskip    
\par The notion of a symplectic geometry is well-known in analytic mechanics through the existence of the Poisson bracket of Hamilton-Lagrange mechanics, {\it i.e.}, for $A$, $B$ functions of the canonical variables $q,p$ on phase space, the Poisson bracket is defined by
$$\{A,B\}_{PB} = \Sigma_{k} \bigl\{{\partial A\over \partial q_k}{\partial B \over \partial p^k} - {\partial B\over \partial q_k}{\partial A \over \partial p^k}\bigr\}. \eqno(2.1)$$
The antisymmetric bilinear form of this expression has the symmetry of
the {\it symplectic group}, associated with the symmetry of the
bilinear form $\xi_i \eta^{ij}\xi_j$, with $i,j= 1,2,\dots 2n$ and
$\eta^{ij}$ an antisymmetric matrix (independent of $\xi$); the
$\{q_k\}$ and $\{p^k\}$ can be considered as the coordinatization of a
symplectic manifold.
\par The coordinatization and canonical mapping of a symplectic manifold [13], to be called a {\it symplectomorphism},  can be constructed by considering two $n$-dimensional manifolds $X_1$ and $X_2$ (to be identified with the target and image spaces of the map) with associated cotangent bundles $M_1= T^*X_1, M_2= T^*X_2$, so that
$$M_1 \times M_2 = T^*X_1 \times T^*X_2 \simeq T^*(X_1 \times X_2). \eqno(2.2)$$
To complete the construction of the symplectomorphism, one defines the involution $\sigma_2$. The action of this involution, in terms of the familiar designation, if  $(x_2,p_2) \in M_2 =T^*X_2$ is a point in $M_2$ (so that $x_2$ is a point in $X_2$ and $p_2$ is a one-form at the point $x_2)$, we define 
$$ \sigma_2 (x_2, p_2) = (x_2, -p_2) \eqno(2.3)$$
We then define 
$$ \sigma = {\rm id}_{M_1} \times \sigma_2, \eqno(2.4)$$
where ${\rm id}_{M_1}$ is the identity map on $M_1$.
\par This construction can be extended to a coordinate patch on $M_2$, enabling the construction of a bilinear form in the tangent space of $M_2$. A vector 
$$ v= v^j {\partial \over \partial u^j}, \eqno(2.5)$$
where, on some coordinate patch on $M_2$ with $u_j = {x_2}^j, j= 1.... n,$ and  $u^j = p_{2, j-n}, j= n+1, .... 2n$, and ${\tilde u} = \sigma_2 u$, 
in the tangent space $TM_2$, gives rise to a one-form; the differential of the map induced by $\sigma_2$ results in the vector (``pushforward''), 
$$ d\sigma_2 (v) = v^j {\partial {\tilde u}^i \over \partial u^j} {\partial \over
\partial {\tilde u}^i}. \eqno(2.6)$$
If $\beta$ is a one-form, the (``pullback'') map ${\sigma_2}^* : T^* M_2 \rightarrow T^* M_2$, defined by 
$$ {\sigma_2}^* \beta (v) = \beta (d\sigma_2 (v)) \eqno(2.7)$$
provides the characteristic antisymmetric form on the symplectic manifold required
for the formulation of Lagrangian mechanics.
\par One then proceeds to define a smooth function $f\in C^\infty(X_1 \times X_2)$; if $df$ is a 
closed $1$-form on $T(X_1 \times X_2) $, call 
$$Y_f= \{((x,y), (df)_{(x,y)}) : (x,y) \in X_1 \times X_2 \}. \eqno(2.8)$$
Then, 
$$ {Y_f}^\sigma = \sigma (Y_f)= \{ ((x,y), d_xf, -d_yf)): (x,y) \in X_1 \times X_2 \}. \eqno(2.9)$$
If ${Y_f}^\sigma$  is a graph of a diffeomorphism $\varphi:M_1 \rightarrow M_2$, then $\varphi$ is a symplectomorphism. Now suppose $\varphi: M_1 - T^* X_1 \rightarrow M_2 = T^*X_2$ is the map
$$ \varphi(x,\xi) = (y,\eta) \eqno(2.10)$$
and ${Y_f}^\sigma$ is its graph, then
$$\eqalign{\xi_i dx^i &= {\partial f \over \partial x^i}dx^i \Rightarrow \xi_i= {\partial f \over \partial x^i}\cr
  \eta_i dy^i &= - {\partial f \over \partial y^i}dy^i \Rightarrow \eta_i= -{\partial f \over \partial y^i}.\cr}\eqno(2.11)$$
We may now attempt to solve $(2.10)$ to obtain
$$ y= y(x,\xi), \eqno(2.12)$$
and then the second of $(2.11)$ to obtain
$$\eta= \eta(x,y(x,\xi)) \equiv \eta(x,\xi) \eqno(2.13)$$
and with this, determine the symplectomorphism    
$$ \varphi(x, \xi) = (y(x,\xi),\eta(x,\xi)). \eqno(2.14)$$
In its application to Hamiltonian mechanics, in the usual notation, let
$$\varphi(q^1,...q^n, p_1,...p_n) = (x^1,...x^n,\pi_1 ...\pi_n)\eqno(2.15)$$
between $M_1=T^* X_1$ and $M_2= T^* X_2$ through the equations
$$\eqalign{p_i &= {\partial f(q,x)\over \partial q^i}\cr
\pi_i &= -{\partial f(q,x)\over \partial x^i}; i= 1,2...n,\cr} \eqno(2.16)$$
where we have denoted the {\it generating function} of the symplectomorphism $\varphi$  by $f$. We remark that the possibility of solving $(2.11)$ locally to obtain $(2.12)$ and $(2.13)$ requires that 
$$ \det \bigl( {\partial^2f(q,x) \over \partial q^i \partial x^j}\bigr) \neq 0. \eqno(2.17)$$ 
\par  The equations $(2.16)$, of the form of the usual canonical transformation derived by adding a total derivative to the Lagrangian in Hamilton-Lagrange mechanics, have been obtained here by a more general and more powerful geometric procedure (the theory of symplectomorphisms), enabling, as we shall see, a simple formulation of the transformation from the standard Hamiltonian form to a geometrical type Hamiltonian.
\bigskip
\noindent{\bf 3. Geodesic Deviation}
\bigskip
\par The principal reason for introducing the canonical
transformation from Hamiltonian form to the geometric form, as we
have pointed out in the introduction, is to make accessible the very
sensitive measure of stability provided by geodesic deviation. In this
section we develop a geometrical formulation of this technique which makes clear the relation between stability in the geometric space and stability in the original Hamiltonian space. 
\par Returning to the geometrical framework defined in Section 2, let ${\bf X}$  be a Hamiltonian vector field in the phase space $M_1$, satisfying
$$ i_{\bf X} \omega = dH, \eqno(3.1)$$
where $\omega$ is the canonical symplectic form on $M_1$.
The integral curves of ${\bf X}$ , obtained by solving Hamilton's equations for $H$, are the trajectories of the Hamilton dynamical system. Since the mapping $\varphi$ to $M_2$ is a symplectomorphism, the pullback by $\varphi$ of the canonical symplectic form ${\tilde\omega}$ on $M_2$ satisfies
$$ \varphi^* {\tilde\omega}= \omega. \eqno(3.2)$$
If $d\varphi:TM_1 \mapsto TM_2$ is the differential of $\varphi$ and we define the vector field ${\bf X}_{geo} = d\varphi({\bf X})$, we have
$$ i_{{\bf X}_{geo}} {\tilde\omega} = dH_{geo}, \eqno(3.3)$$
so that ${\bf X}_{geo}$ is a Hamiltonian vector field in $TM_2$ with respect to the Hamiltonian function $H_{geo}$; the integral curves for ${\bf X}_{geo}$ correspond to geodesics in $M$. We shall refer to such integral curves of ${\bf X}_{geo}$ as $M_2$ geodesics, or {\it cotangent bundle geodesics}.
\par Let $\gamma \subset M_1$ be a trajectory in phase space of the original dynamical system. Then, $\gamma^\varphi = \varphi(\gamma)$ is an $M_2$ geodesic. If ${\tilde \pi}: M_2 \rightarrow M$ is the projection of the cotangent bundle $M_2 =T^*M$ on the base manifold $M$, then ${\tilde \pi} (\gamma^\varphi)$ is a geodesic in $M$. For $G$ the map of the tangent bundle $M_3 = TM$ to the cotangent bundle $M_2$, we apply the inverse map $G^{-1}: M_2 \mapsto M_3$, the tangent bundle for $M$, {\it i.e.} $(x, {\bf v})$, where ${\bf v} \in T_x M$ ($x$ is a point in $M$), to $\gamma^\varphi$, we obtain an $M_3$  (or tangent bundle) geodesic
$$\gamma^Q= G^{-1}(\gamma^\varphi)  = (G^{-1}\circ \varphi)\gamma = Q(\gamma).\eqno(3.4)$$
If now $\pi:M_3 \mapsto M$ is the projection of the tangent bundle on the base manifold $M$, then $\pi (\gamma^Q ) = {\tilde \pi} (\gamma^\varphi )$ is a geodesic in $M$. This establishes the equivalence of trajectories in the original Hamiltonian space with geodesics in the geometric space.
\par Let $u_0 \in M_1$ be a point in phase space and let $\gamma_0\subset M_1$ be the curve given by $\gamma_0(t) = \phi_t(u_0)$, where $\phi_t$ is the flow in the phase space $M_1$ of the Hamiltonian dynamical system generated by$H$, {\it i.e.}, $\gamma_0$ is a trajectory of the system such that $\gamma_0(0) = u_0$. Let ${\tilde W}^{2n-1} \subset M_1$ be a surface of section at $u_0$, {\it i.e.} a hypersurface in $M_1$ transverse to the trajectories of the dynamical system and defined in some open neighborhood of $u_0$. Let $E_0 \subset M_1$ be an equal energy hypersurface passing through a point $p_0 \in E_0$, for which $dH =0$ on $E_0$, and let $W = W^{2n-1} \cap E_0$.  Then $W$ is a $2n-2$ dimensional submanifold of $M_1$ such that the Hamiltonian  $H$ has the same value at all points $u \in W$ and such that the trajectories of  the dynamical system are transverse to $W$ at all points of intersection. Now, let $u$ be an arbitrary point in $W$; then it is a base point of a trajectory $\gamma_u$ given by $\gamma_u (t) = \phi_t (u)$. In a time interval $0 \leq t \leq T (T> 0)$ we define a submanifold $N_{u_0} \subset M_1$ by
$$ N_{u_0} = \{\phi_t (u) : \forall u\in W, \forall t \in [0,T]\}. \eqno(3.5)$$
Then, $N_{u_0}$ is parametrized by $(u,t)$, for $u\in W, t \in [0,T] $. and consists of trajectories of the dynamical system corresponding to all initial points $u\in W$. Now apply the mapping $Q$ to obtain a submanifold  ${N_{u_0}}^Q \subset M_1$ according to
$$ {N_{u_0}}^Q = Q(N_{u_0})= \{Q[\phi_t(u)]: \forall u\in W, \forall t\in [0,T]\}. \eqno(3.6)$$ 
Again, by construction, ${N_{u_0}}^Q$ is parametrized by  $(u,t)$, for $u\in W, t \in [0,T] $.  For each $u\in W$, the curve ${\gamma_u}^Q = Q(\gamma_u)$ is an $M_3$ geodesic curve given by ${\gamma_u}^Q (t)= Q[\phi_t(u)]$ and ${N_{u_0}}^Q$ consists of all such geodesic curves corresponding to all possible initial points $u\in W$. In particular, ${\gamma_0}^Q = Q(\gamma_0)$ is the $M_3$ geodesic corresponding to to the trajectory $\gamma_0$ of the original dynamical system.        
\par To calculate geodesic deviation, we now consider variations of such trajectories.  Let $\gamma_{var}\subset W$ be a curve parametrized by a parameter $\alpha$ and based at the point $u_0 \in W$. For some interval $I\subset {\bf R}$, with $0\in I$, $\gamma_{var}$ is given by a smooth function  $u(\alpha)\in W, \forall \alpha \in I$ and $u(0) = u_0$. The curve $\gamma_{var}$ corresponds to a two dimensional surface $S_{var} (\gamma_{var})\subset N_{u_0}$ through the definition
$$ S_{var} (\gamma_{var}) = \{\phi_t(u(\alpha)): \alpha \in I, t\in [0,T] \} \eqno(3.7)$$
By construction, $(t,\alpha), t\in [0,T], \alpha \in I$ are coordinates on  
$S_{var} (\gamma_{var})$, the {\it variational surface } of $\gamma_0$ corresponding to $\gamma_{var}$. Each such curve $\gamma_{U(\alpha)}$, given by  $ \gamma_{var}(t) = \phi_t(u(\alpha), t\in [0,T]$, is a trajectory of the original Hamiltonian system.  Furthermore, $\gamma_{var}$ is carried by the flow $\phi_t$ to a variation curve ${\gamma^t}_{var}$ at time $t$ defined by ${\gamma^t}_{var} = \phi_t(\gamma_{var})$, given explicitly by the function  $\phi_t(\alpha) = \phi_t(u(\alpha))$, where $u(\alpha)$ is the function defining ${\gamma^t}_{var}$. Applying the mapping $Q$ to $  S_{var} (\gamma_{var})$, we obtain an $n-1$ dimensional surface in $M_3$ (two dimensional surface in a three dimensional problem)

$$\eqalign{ {S^Q}_{var} (\gamma_{var}) &= Q[ S_{var} (\gamma_{var})]\cr
&= \bigl\{ {\gamma^Q}_{u(\alpha)} : \alpha \in I \bigr\} \cr &=\{Q(\gamma_{u(\alpha)} ): \alpha \in I\}\cr &= \{Q[\phi_t (u(\alpha))]: \alpha \in I, t\in [0,T] \}, \cr} \eqno(3.8)$$
where
$$ {\gamma^Q}_{u(\alpha)}= Q(\gamma_{u(\alpha)}) =  Q[\phi_t(u(\alpha))]. \eqno(3.9)$$
Note that $(t,\alpha), \ \ \ t\in [0,T], \alpha \in I $ are coordinates on ${S^Q}_{var} (\gamma_{var})$, and  that, since each curve $\gamma_{u(\alpha)}$ is a trajectory of the original dynamical system,  ${\gamma^Q}_{u(\alpha)}$ is an $M_3$ geodesic. Therefore, ${S^Q}_{var} (\gamma_{var})$ is a  surface of variation for ${\gamma_0}^Q$ consisting of $M_3$ geodesics. Furthermore, ${\gamma_{var}}^{Q,t} = Q({\gamma_{var}}^t = [\phi_ t(\gamma_{var})] $ is  the variation at time $t$ in $ {S^Q}_{var} (\gamma_{var})$ corresponding to the variation curve ${\gamma_{var}}^t \subset  S_{var} (\gamma_{var})$.  A parametrization of ${\gamma_{var}}^{Q,t}$ is provided by  the function ${\gamma_{var}}^{Q,t}(\alpha) = {\gamma_{u(\alpha)}}^Q (t),\ \  \alpha \in I$, with $t$ constant.
\par We now wish to investigate the deviation of nearby trajectories of the original Hamiltonian system by considering the deviation of the corresponding geodesics in $M_3$. We quantify the deviation of nearby trajectories from the base trajectory $\gamma_0$ in $N_{u_0}$, {\it i.e.}, on the variational surface $S_{var}(\gamma_{var})$, by studying the evolution along $\gamma_0$ of the tangent vector to the variation curve ${\gamma_{var}}^t $ . The tangent vector, which we call the {\it phase space trajectory deviation vector } is formally given by
$$ {\bf V}_{trj} (t) = \bigl[ {\partial \over \partial \alpha} {\gamma_{var}}^t (\alpha) \bigr]|_{\alpha =0} = \bigl[ {\partial \over \partial \alpha} \phi_t(u(\alpha)  \bigr]|_{\alpha =0} \ \ \ \ ,{\bf V}_{trj} (t)\in TM_1. \eqno(3.10)$$

The deviation vector${\bf V}_{trj} (t)$ is mapped by the differential of the mapping $Q$ into a deviation vector in $TM_3$, formally given by
$$ \eqalign{{\bf J}_{dev}(t) & = \bigl[ {\partial \over \partial \alpha} {\gamma_{var}}^{Q,t}(\alpha)\bigr]_{\alpha =0}
= \bigl[ {\partial \over \partial \alpha} {\gamma_{u(\alpha)} }^Q (t)\bigr]_{\alpha =0}  \cr
&= \bigl[{\partial \over \partial \alpha} Q[\phi_t(u(\alpha))]\bigr]_{\alpha =0}\cr
&= dQ\bigl(  \bigl[{\partial \over \partial \alpha} \phi_t(u(\alpha))\bigr]_{\alpha =0} \bigr)
= dQ({\bf V}_{trj} (t)),  \ \ \ {\bf J}_{dev}(t)\in TM_3. \cr} \eqno(3.11)$$
where $dQ: TM_1 \mapsto TM_3$ is the differential of the map $Q$.
\par In order to obtain a more explicit expression for ${\bf J}_{dev}(t)$ we will need a more explicit expression for the points in ${N^Q}_{u_0} \subset M_3$ and, in particular, points in
$ {S^Q}_{var} (\gamma_{var})$. Recall the fact that  $(t,\alpha),t \in [0,T], \alpha \in I$ serve as coordinates in ${N^Q}_{u_0}$. The point corresponding to the pair $(t,\alpha)$ is  ${\gamma_{u(\alpha)} }^Q (t)= Q[\phi_t(u(\alpha))] = (x(x,\alpha), {\bf T}(t,\alpha)),$ where $x(t,\alpha) = \pi({\gamma^Q}_{u,\alpha}) \in M$ is a point on the geodesic $\pi\bigl({\gamma_{u(\alpha)} }^Q (t)\bigr)$ at the point $x(t,\alpha)$.  Since ${\bf T}(t,\alpha)$ forms a vector field defined on $\pi ({N^Q}_{u_0})$ and, in particular, along the geodesic curve ${\gamma^{Q,t}}_{var}$, its $\alpha$ derivative is given by the covariant derivative ${\nabla {\bf T}(t,\alpha)\over \partial \alpha}$. Then, we find that
$${\bf J}_{dev} (t) = \bigl[ {\partial \over \partial \alpha} Q[\phi_t(u(\alpha))\bigr]_{\alpha =0} = \bigl( {\partial x(t,\alpha)\over \partial \alpha}) |_{\alpha =0} , {\nabla {\bf T}(t,\alpha)\over \partial \alpha}|_{\alpha =0} \bigr)^T
\eqno(3.12)$$
Note that ${\bf J}_{dev} (t) \in T_{x(t,0)}M \oplus T_{x(t,0)}M = T M_3.$
\par The standard definition of the geodesic deviation vector for geodesics in $M$ is
$${\bf J} (t)=  \bigl({\partial x(t,\alpha)\over \partial \alpha}\bigr)|_{\alpha =0},\ \ \ {\bf J} (t)\in T_{x(t,0)}M. \eqno(3.13)$$ 
According to Theorem 10 of Frankel [16],
$$ {\nabla {\bf J}(t) \over \partial t} = \bigl( {\nabla {\bf T}(t,\alpha)\over \partial \alpha}|_{\alpha =0}, \eqno(3.14)$$
so that
$$ {\bf J}_{dev}(t) = \bigl({\bf J}(t), {\nabla {\bf J}(t) \over \partial t} \bigr)^T , \eqno(3.15)$$
where $t$ is the affine parameter parametrizing ${\gamma_0}^Q$. 
\par The equation of evolution of ${\bf J}_{dev}(t)$, {\it i.e.} the
{\it dynamical system representation} of the geodesic deviation
equation, has been studied in ref [14].
\par   Let ${\bf X,Y,Z}\in T_pM $ be ($n$=dimensional) vectors and let $R_p({\bf X,Y}): T_pM\mapsto T_pM$ be the curvature transformation at the point $p\in M$ {\it i.e.}, the linear transformation with matrix elements ${[R_p({\bf X,Y})]_j}^i = {R^i}_{jk\ell} X^i Y^j$ so that
$$ R_p({\bf X,Y}){\bf Z} = ({R^i}_{jk\ell} X^k Y^\ell Z^j)\partial_i, 
\eqno(3.16)$$
where $\partial_ i$ are coordinate vectors at $p$ and ( $X^k,Y^k, Z^k, 1\leq k \leq n$ are the components of ${\bf X,Y,Z}$ with respect to the basis ${\{ \partial_k\}_{k=1}}^n$).  The quantities ${R^i}_{jk\ell}$ are the components of the Riemann curvature tensor at the point $p$.
\par Furthermore, if $<\cdot, \cdot >_{T_pM}$ denotes the inner product defined on $T_pM$ with the metric $g(\cdot,\cdot)$ on $M$, then for ${\bf W} \in T_pM$ we have
$$<R_p({\bf X,Y}){\bf Z},{\bf W}>_{T_pM} =  {R^i}_{jk\ell} X^k Y^\ell Z^j W_i, \eqno(3.17)$$
where $W_i = g_{ij}W^j$.  For the geodesic ${\gamma_0}^Q \in M$, given in terms of the function ${\gamma_0}^Q (t) = Q[\phi_t(u_0)]$, using the above  notation for the curvature transformation, the geodesic deviation equation along ${\gamma_0}^Q$ is
$$ {\nabla^2 {\bf j}(t) \over dt^2} + R_{{\gamma_0}^Q (t)} ({\bf J}(t), {\bf T}(t))({\bf T}(t)) =0. \eqno(3.18)$$ 
where ${\bf J}(t)$ is the geodesic deviation vector defined above, ${\bf T}(t) \equiv {\bf T}_{{\gamma_0}^Q (t)}$ is the tangent vector to ${\gamma_0}^Q$ at the point ${\gamma_0}^Q (t)$ and $R_{{\gamma_0}^Q (t)}$ is the curvature tensor at the point ${\gamma_0}^Q (t)$. The dynamical system representation of the geodesic deviation equation corresponds to putting $(9.16)$  into the form
$$ {\nabla \over dt}\left(\matrix{{\bf J}(t) \cr{\nabla {\bf J}(t) \over dt}\cr}\right)= 
\left(\matrix{0&I\cr - R_{{\gamma_0}^Q (t)}(\cdot, {\bf T}(t)) {\bf T}(t)&0 \cr}\right) \left(\matrix {{\bf J}(t) \cr {\nabla {\bf J}(t) \over dt}\cr}\right). \eqno(3.19)$$
Denoting
$$ {\hat R}_{{\gamma_0}^Q (t)} = \left(\matrix{0&I\cr - R_{{\gamma_0}^Q (t)}(\cdot, {\bf T}(t)) {\bf T}(t)&0 \cr}\right)
\eqno(3.20)$$
and using $(3.15)$, we may write $(3.19)$ in the shorter form
$$ {\nabla {\bf J}_{dev} \over dt} = {\hat R}_{{\gamma_0}^Q (t)}{\bf J}_{dev} , \eqno(3.21)$$
The behavior of  the solution $ {\bf J}_{dev}$ of the equation $(3.21)$ determines the deviation properties of geodesics near ${\gamma_0}^Q$ as a function of $t$ and, through the relation ${\bf V}_{trj} (t)= dQ^{-1} ( {\bf J}_{dev}(t)) $ obtained from $(3.11)$, also the deviation of trajectories of the original dynamical system near $\gamma_0$ over time. The deviation of trajectories of the original system near $\gamma_0$ is therefore governed by the curvature transformation
$R_{{\gamma_0}^Q (\cdot)}$ along the geodesic ${\gamma_0}^Q (\cdot)$.  

\bigskip
\noindent{\bf 4. Formulation of the Algorithm}
\bigskip
\par The purpose of the canonical transformation we have discussed above is to construct a Hamiltonian of the geometrical form $(1.2)$ by means of a canonical transformation from a Hamiltonian of the form $(1.1)$. As above, we label the coordinates and momenta of the image space by $\{x_i\}$ and $\{\pi^i\}$ (we do not require that $p^i$ and $\pi^i$ are necessarily simply related for all $t$ here; the equivalence of the dynamics is assured by the canonical nature of the transformation).  We must therefore find the generating function $f(q,x)$ and the metric $g_{ij}(x)$ from the statement 
$$ { p^2 \over 2m} + V(q)=  {1\over 2m}g_{ij}(x)\pi^i \pi^j \eqno(4.1)$$
Substituting $(2.16)$ for the momenta, the problem is to solve (note that the left hand side treats the indices as Euclidean since it does not carry the local coordinate transformations available to the geometric form on the right hand side)  
$$ V(q) +{1\over 2m } {\partial f(q,x) \over \partial q_i}{\partial f(q,x) \over \partial q_i}
=  {1\over 2m }g_{ij}(x){\partial f(q,x) \over \partial x_i}{\partial f(q,x) \over \partial x_j} \eqno(4.2)$$
\par Assuming analyticity in the neighborhood of the origin of the coordinates $\{q\}$ , and in the potential term $V(q)$, one can write a power series expansion of the generating function and the potential, and identify the resulting powers of $q_i, q_j...$ and their products. This procedure provides an effective recursive algorithm for a system of nonlinear first order equations in the expansion coefficients since the powers of $q$ on the right hand side occurring in the expansion of $f(q,x)$ are higher by one order that the expansions on the left hand side, which contain derivatives with respect to $q$. Assuming analyticity in $\{x\}$ as well near the origin (as for Riemann normal coordinates), one can find a recursion relation for the resulting coefficients.
\par For example, in two dimensions, one may expand, into some radius of convergence,
$$ f(q^1, q^2, x^1,x^2) = \Sigma_{k,\ell =0}^\infty C_{k,\ell} (x^1,x^2) (q^1)^k (q^2)^\ell \eqno(4.3)$$
and expand $V(q^1, q^2)$ in power series
$$ V(q^1,q^2) = \Sigma_{k,\ell =0}^\infty v_{k,\ell}(q^1)^k q^2)^\ell \eqno(4.4)$$
Substituting into the relation $(4.2)$ (in two dimensional form), and equating coefficients of powers of $q^1$ and $q^2$, one finds the following recursion relations:
$$\eqalign{v_{k,\ell} +\Sigma_{k=0}^m \Sigma_{\ell =0}^n &\bigl[(k+1)(m-k+1)C_{(k+1),\ell}(x^1,x^2) C_{(m-k+1),(n-1)}(x^1,x^2) + (\ell+1)(n-\ell+1)C_{k,(\ell +1)}(x^1,x^2) C_{(m-k),(n-\ell +1)} (x^1,x^2) + 2v_{n,m}\bigr]\cr
&=\Sigma_{k=0}^m \Sigma_{\ell =0}^n\bigl[ g^{11}(x^1,x^2){\partial C_{k,\ell}\over \partial x^1}(x^1,x^2) {\partial C_{m-k,n-1}\over \partial x^1}(x^1,x^2) \cr
&+2g^{12}(x^1,x^2){\partial C_{k,\ell}\over \partial x^1}(x^1,x^2) {\partial C_{m-k,n-1}\over \partial x^2}(x^1,x^2)\cr
&+ g^{22}(x^1,x^2){\partial C_{k,\ell}\over \partial x^2}(x^1,x^2) {\partial C_{m-k,n-1}\over \partial x^2}(x^1,x^2)\bigr]\cr} \eqno(4.5)$$
\par The solution of this system of equations, for a given potential $V$ requires, even in two dimensions, significant computational power. Our initial investigations indicate reasonable behavior, with strong indications of convergence, for some simple cases.
\par Although the physically interesting cases are in two or more
dimensions, where curvature generated by the geometric Hamiltonian
plays an important role in the formation of geodesic curves and for many
practical problems, we shall describe the general structure of the
calculation in one dimension below as well as to give a convergence
proof for this case, which, it appears, can be extended to arbitrary
dimension. Some basic properties of the higher dimensional structure
are discussed below as well, but a full development of the algorithm in
higher dimensions and applications will be treated in succeeding publications. 
\bigskip
\noindent{\bf 5. One dimensional study}
\bigskip
\par In one dimension, Eq. $(4.2)$ becomes
$$ V(q) +{1\over 2m } \bigl({\partial f(q,x) \over \partial q}\bigr)^2
=  {1\over 2m }g(x)\bigl({\partial f(q,x) \over \partial x}\bigr)^2 \eqno(5.1)$$
\par The recursion relation for the one dimensional case for
$$ \eqalign{f&= \Sigma q^\ell C_\ell (x)\cr
V(q) &= \Sigma_\ell V^{(\ell)} q^\ell \cr} \eqno(5.2)$$
becomes
$$\eqalign{\Sigma_{m=0}^\ell\{ (\ell + 1 -m)(m+1) &C_{\ell +1-m} C_{m+1} \cr
&-g(x)C'_{\ell - m} C'_m \} + V^{(\ell)} = 0\cr} \eqno(5.3)$$
Now, taking
$$ \eqalign{ C_\ell(x) &= \Sigma_0^\infty b_{\ell m} x^m \cr
g(x) &= \Sigma_0^\infty g_n x^n 
\cr} \eqno(5.4)$$
we find (for coefficients of $x^r$)
\par $r=0$:
$$\eqalign{\Sigma_{m=0}^\ell \bigl\{(\ell + 1 -m)(m+1)b_{\ell + 1 -m,0} b_{m+1, 0}\cr
&-g_0 b_{\ell -m,1} b_{m,1}\bigr\} + V^{(\ell)} =0\cr} \eqno(5.5)$$
and for
\par $r \geq 1$ :
$$\eqalign{\Sigma_{m=0, 0\leq p\leq r}^\ell &(\ell + 1 -m)(m+1)b_{\ell + 1 -m,p} b_{m+1, r-p}\cr
&-\Sigma_{n, 1\leq p\leq r+1}^\ell g_n b_{\ell -m,p} b_{m,r-n-p+2}\cr
&\times p(r-n-p+2)=0. \cr} \eqno(5.6)$$
Note that for the case $r \geq 1$, the potential does enter explicitly since it has no $x$ dependence.  The relations $(5.5)$ and $(5.6)$ provide the basis  for a systematic recursion.     
\par One can easily work out several terms to see  how the algorithm
develops. It is clear that it is iteratively closed, but it is
difficult to draw detailed conclusions on the solutions without
extensive computations, as well as specification of potential models. 
\par We give in the next section a proof, however, for one dimension,
that, with some reasonable assumptions, such a computation converges. The method of proof can be generalized to $n$ dimensions.
\bigskip
\noindent{\bf 6. Convergence of the algorithm in one dimension}
\bigskip
Now, in $(5.3)$, define
$$ D_m = m C_m, \eqno(6.1)$$
and note that the first term in $(5.3)$ can then be written as
$$ \Sigma_{m=0}^\ell D_{\ell +1 -m} D_{m+1}= \Sigma_{m=1}^{\ell +1} D_n {A^{(\ell)}}_{nm}D_m, \eqno(6.2)$$
where symmetric the matrices ${A^{(\ell)}}_{nm}$  consist of completely skew diagonal $1$'s, a reflection of the combinatorial origin of the coefficients. The trace is zero for even and unity for odd $\ell$'s, and the eigenvalues are $\pm 1$. They can occur in any order, but the orthogonal matrices that diagonalize $A^{(\ell)}$ may be constructed so that that the  eigenvalues slternate (this is convenient for our proof of convergence but not necessary). Let us call these orthogonal matrices ${u^{(\ell)}}_{nm}$ and represent the ``vectors'' $D_m$ in terms of the eigenvectors $d_n^\ell$ as
$$ D_m = \Sigma_{n=1}^{\ell +1} {u^{(\ell)}}_{mn}{d_n}^\ell, \eqno(6.3)$$
where
$$  \Sigma_{n=1}^{\ell+1}{u^{(\ell)}}_{mn}{u^{(\ell)}}_{m'n}= \delta_{m m'}. \eqno(6.4)$$
We then obtain
$$\eqalign{ \Sigma_{m=0}^\ell D_{\ell +1 -m} D_{m+1} &= \Sigma_{m=1}^{\ell +1} D_n {A^{(\ell)}}_{nm}D_m\cr
&= \Sigma_{m=1}^{\ell +1} {\lambda^{(\ell)}}_m {({d_m}^\ell)}^2.\cr} \eqno(6.5)$$
Now, consider the sum in the second term of $(5.3)$:
$$\Sigma_{m=0}^\ell  C'_{\ell-m}(x) C'_m (x) = \Sigma_{m=0}^\ell C' _m {B^{(\ell)}}_{mn} C'_n , \eqno(6.6)$$
where ${B^{(\ell)}}_{mn} = {A^{(\ell)}}_{m+1,n+1}$, the same set of matrices as $A^{(\ell)}$, occurring here with indices $1,....\ell+1$ as well. By shifting the indices in the vectors $C'_n$ by unity, one obtains the same structure as for the left hand side, {\it i.e.}
for $m=0,...\ell$, and $f$ the eigenvectors constructed from $C'$,
$$ C'_{m-1} = \Sigma_{n=1}^{\ell +1} {u^{(\ell)}}_{mn}{f_n}^\ell. \eqno(6.7)$$
We then have 
$$ \Sigma_{m=0}^\ell C'_m {B^{(\ell)}}_{mn} C'_n = \Sigma_{m=1}^{\ell +1} {\lambda^{(\ell)}}_m  {({f_m}^\ell)}^2 \eqno(6.8)$$
so that our condition for a solution to the equations $(5.3)$ becomes
$$ V^{(\ell)} + \Sigma_{m=1}^{\ell +1} {\lambda^{(\ell)}}_m [({d_m}^\ell)^2 - g(x)({f_m}^\ell)^2]=0. \eqno(6.9)$$
\par We now study the convergence of the $d$ and $f$ sums as $\ell \rightarrow \infty$.
Inverting $(6.3)$ and $(6.7)$, we obtain
 $$ {d_m}^\ell = \Sigma_{n=1}^{\ell+1} n C_n {u^{(\ell)}}_{nm} \eqno(6.10)$$
and 
  $$ {f_m}^\ell = \Sigma_{n=1}^{\ell+1}  C'_{n-1} {u^{(\ell)}}_{nm}. \eqno(6.11)$$
Since $ {u^{(\ell)}}_{nm}$ is an orthogonal matrix, it follows that
$$ \Sigma_{n=1}^{\ell+1} ({f_m}^\ell)^2 = \Sigma_{n=1}^{\ell+1} {C'_{n-1}}^2  \eqno(6.12)$$
 and
$$ \Sigma_{n=1}^{\ell+1} {({d_m}^\ell)}^2 = \Sigma_{n=1}^{\ell+1} n^2 {C_n}^2  \eqno(6.13)$$
 It is sufficient to argue that the sequences in these sums are decreasing. The alternating (due to the ${\lambda_m}^\ell$) series appearing in $(6.9)$ then converges. 
\par We first remark that the generating function $f(q,X)$ is $C^\infty$ in both variables, so that all orders of derivative with respect to $q$ exist.   We seek solutions that can be represented as power series in $q$. Suppose that  this series converges for all values of $q <q_0 (x)$ (the radius of convergence can depend on $x$), and call $D_\epsilon$ the domain of $x$ such that 
 $|q_0(x)|\geq\epsilon>0$, The ratio test prescribes that, for each
 such $x$,
$$ |{C_{\ell +1} \over C_\ell}| < {1 \over |q_0(x)|}\eqno(6.14)$$
\par The series $(4.2)$ corresponds to the Taylor expansion 
 $$ f(q,x) = \Sigma_0^\infty { 1 \over \ell!\, }f^\ell, \eqno(6.15)$$
where 
$$ f^\ell = { \partial^\ell f \over \partial q^\ell}; \eqno(6.16)$$
The ratio condition then becomes
$$ |{f^{\ell +1} \over f^\ell}| < |{\ell +1 \over q_0}|. \eqno(6.17)$$
If the derivatives do not grow faster than linearly, this condition should be satisfied for sufficiently large $\ell$. Taking $|q_0| =  \epsilon $, the convergence would be uniform in $D_\epsilon$.
\par Now, consider the decreasing property. As for any  series depending on a dimensional variable, we may scale the dimension, for $|q_0| >0$, so that $|q_0(x)|>1 $ for all $x \in D_\epsilon$ (the ratio ${C_{\ell +1} / C_\ell}$  scales with $1/q$ as well). This choice of scale is adequate for all $x \in D_\epsilon$ for a scale such that $\epsilon >1$. Then, uniformly, the $|C_\ell(x)|$ forms a decreasing sequence, leading to convegence of the  $d$ series in $(6.9)$ (the factor $m$ in $(6.1)$ does not affect the convergence for large $m$).  A similar argument can be followed for the $f$ series following the convergence of the series in $q$ for $\partial f(q,x)/\partial x$.
\par This completes our proof of convergence.
\par As remarked in the introduction, the nonliear expansions can be studied by means iof Fourier series representations in terms of (upper half place) analytic functions (see, for example [15]), which may provide useful approximation techniques in specific cases.  This study will appear in a later publication.
\bigskip
\noindent{\bf 7. Shift of Origin for Expansion}
\bigskip
\par We now return to arbitrary dimension. The algorithm proposed in
Section 3 contains an expansion of the potential function $V(q)$ around
some point $q=0$; for a polynomial potential or some other entire
function, there would be no question of convergence of this
expansion, but the algorithm itself may have only a finite domain of
convergence. To extend the range of the resulting functions, it would
then be necessary to carry out the expansions around some new origin
at,{\it e.g.}, $q=q_0$.    
\par Therefore, let us  now consider expanding $V(q)$ around  $q_0$,
and carry out the same procedure. We then rewrite $(3.2)$ for the modified problem with a new potential function
 $$ V'(q) = V(q+q_0) \eqno(7.1)$$
as
$$ V'(q) + \delta_{ij}{ 1 \over 2m} {\partial {\tilde f}(q,x') \over \partial q_i}{\partial {\tilde f}(q,x') \over \partial q_j}= g_{ij}(x')
{\partial {\tilde f(q,x')} \over \partial x'_i}{\partial {\tilde f}(q,x') \over \partial x'_j},\eqno(7.2)$$
where we observe that the solutions ${\tilde f}(q,x')$ and the
manifold which we label $x'$ will be different from $f(q,x)$ on the manifold $x$ since the
potential function $V'(q)$ is different; however, the variable $q$ on
the original space is still designated by $q$ since it is the argument
of $V'(q)$.
\par The assumptions underlying $(7.2)$ imply that in the generating function ${\tilde
f}(q,x')$, $q$ and $x'$ are independant variables; we may then proceed by
recognizing that, as a result of the solution algorithm, $x'$ can only
be a function of $x$ in the mapping $q,x \rightarrow q,x'$.
\par We can now use the chain rule of derivatives for the right hand side and  consider
 ${\tilde f}(q,x')$ as a function of $q,x$, at least locally under this map. Calling this function $h(q+ q_0,x)$, we can rewrite $(7.2)$ as
$$V'(q) +  \delta_{ij}{ 1 \over 2m} {\partial h(q+q_0,x) \over \partial q_i}{\partial h(q+q_0,x) \over \partial q_j}={\tilde g}_{ij}(x)
{\partial h(q+q_0,x) \over \partial x_i}{\partial h(q+q_0,x) \over \partial x_j},\eqno(7.3)$$
 where 
$${\tilde g}_{ij}(x)= g_{k\ell}(x'){\partial x_i \over \partial x'_k}{\partial x_j \over \partial x'_\ell}.  \eqno(7.4)$$
Replacing as a change of variables $q+q_0 \rightarrow q$, $V'(q)$ becomes $V(q)$, and 
$(7.3)$ becomes
$$V(q) +  \delta_{ij}{ 1 \over 2m} {\partial h(q,x) \over \partial q_i}{\partial h(q,x) \over \partial q_j}={\tilde g}_{ij}(x)
{\partial h(q,x) \over \partial x_i}{\partial h(q,x) \over \partial x_j},\eqno(7.5)$$
\bigskip
\par Since this equation has a solution (among others) of the form for which 
$$ {\tilde g}_{ij}(x)= g_{ij}(x), \eqno(7.7)$$
by applying the same algorithm, we may choose this solution with the consequence that
$$ g_{k\ell}(x'){\partial x_i \over \partial x'_k}{\partial x_j \over \partial x'_\ell}= g_{ij}(x).  \eqno(7.8)$$
With this choice we may follow shifts from $q \rightarrow
q_0 \rightarrow q_1....$ within the domains of convergence choosing
the same algorithm for solution at every step, building a set of
overlapping neighborhoods which consruct a manifold, on which
covariance is maintained through the canonical transformation.
 \bigskip

\noindent{\bf 8. Change in generating function induced by diffeomorphisms in the geometric space}
\bigskip
\par The structure of the image space has the property of supporting local diffeomorphisms. However, our construction concerns a mapping from the the coordinates $\{q,p\}$ to $\{x,\pi\}$; therefore a diffeomorphism of the latter set of variables necessarily involves a change in the generating fumction of the transformation.   
\par In this section, we calculate the effect of an infinitesimal
coordinate transformation on the geometrical space, holding the
Hamiltonian variables $\{q,p\}$ unchanged, on the generating function of the canonical transformation, {\it i.e.}, 
$ f \rightarrow {\tilde f}$. 
\par On the original choice of coordinates, for which
$$ \eqalign{ p_i &= {\partial f(q, x)\over \partial q_i} \cr
\pi_i &= - {\partial f(q, x)\over \partial x_i} \cr} \eqno(8.1)$$
we now consider a new mapping from $q,p$ to $x',\pi'$ differing infinitesimally from $x,\pi$ according to
$$ x'_i = x_i + \lambda_i(x), \eqno(8.2)$$
where $ \lambda_i(x)$ is small.
\par After this mapping, we can write
$$ \eqalign{ p_i &= {\partial {\tilde f}(q, x')\over \partial q_i} \cr
\pi'_i &= - {\partial {\tilde f}(q, x')\over \partial x'_i} \cr} \eqno(8.3)$$
\par To study ${\tilde f}(q,x')$, let us define
$$ g^i (q,x') = {\partial {\tilde f}(q, x')\over \partial x'_i}= -\pi'_i . \eqno(8.4)$$
Then, 
$$ g^i (q, x+ \lambda) \cong {\partial {\tilde f}(q, x)\over \partial x_i}+  {\partial^2 {\tilde f}(q, x)\over \partial x_i\partial x_j}\lambda_j(x) \eqno(8.5)$$
so that 
$$ -\pi'^i \cong  {\partial {\tilde f}(q, x)\over \partial x_i}+  {\partial^2 {\tilde f}(q, x)\over \partial x_i\partial x_j}\lambda_j(x). \eqno(8.6)$$
This result could have been obtained directly from $(8.3)$ but it is perhaps helpful to define the function $g^i (q,x')$ to clarify the computation.   
\par We now impose invariance of
$$ \pi'^i dx'_i = \pi^i dx_i, \eqno(8.7)$$
which leads, through the Hamilton-Lagrange construction, to invariance of the Hamiltonian. We now write out
 $$\eqalign{ -\pi'^i dx'_i &\cong \bigl[ {\partial {\tilde f}(q, x)\over \partial x_i}+  {\partial^2 {\tilde f}(q, x)\over \partial x_i\partial x_j}\lambda_j(x)\bigr]\cr
&\times\bigl[ dx_i + {\partial \lambda_i \over \partial x_k} dx_k \bigr] \cr
 &=  {\partial {\tilde f}(q, x)\over \partial x_i}dx_i+  {\partial^2 {\tilde f}(q, x)\over \partial x_i\partial x_j}\lambda_j(x)dx_i \cr
&+ {\partial {\tilde f}(q, x)\over \partial x_i}{\partial \lambda_i \over \partial x_k} dx_k + {\partial^2 {\tilde f}(q, x)\over \partial x_i\partial x_j}\lambda_j(x){\partial \lambda_i \over \partial x_k} dx_k\cr
&= -\pi^i dx_i. \cr} \eqno(8.8)$$
Therefore, to order $\lambda dx $, 
$$\eqalign{ dx_i  {\partial f(q, x)\over \partial x_i}&= dx_i \bigl\{   {\partial {\tilde f}(q, x)\over \partial x_i}+  {\partial^2 {\tilde f}(q, x)\over \partial x_i\partial x_j}\lambda_j(x) \cr
&+ {\partial {\tilde f}(q, x)\over \partial x_k}{\partial \lambda_k \over \partial x_i}\bigr\} \cr
&= dx_i \bigl\{ {\partial {\tilde f}(q, x)\over \partial x_i} + {\partial \over \partial x_i} \bigl[{\partial {\tilde f}(q, x)\over \partial x_k}\lambda_k \bigr] \bigr\}, \cr } \eqno(8.9)$$
so that
$$dx_i  {\partial f(q, x)\over \partial x_i}= dx_i {\partial \over \partial x_i}\bigl[ {\tilde f}(q,x) + \lambda_k {\partial {\tilde f}(q, x)\over \partial x_k}\bigr] \eqno(8.10) $$
 If we write (say, integrate up to some $x_i$)
$$ f(q,x) = {\tilde f}(q,x) + \lambda_k {\partial {\tilde f}(q, x)\over \partial x_k}, \eqno(8.11)$$
we may approximately invert to get
$$ {\tilde f}(q,x)\cong {f(q,x)-  \lambda_k} {\partial  f(q, x)\over \partial x_k} \eqno(8.12)$$
This corresponds to a conformal-like local transformation. The algebra of such generators is
$$[ {\lambda_i}^a {\partial \over \partial x_i}, {\lambda_j}^b {\partial \over \partial x_j}] = \bigl( {\lambda_i}^a {\partial {\lambda_j}^b \over \partial x_i} - {\lambda_i}^b {\partial {\lambda_j}^a \over \partial x_i} \bigr) {\partial \over \partial x_j} \eqno(8.13)$$
Thus the algebra is of a conformal type, but the coefficients may run on, so that the group may not be finite dimensional.
\smallskip
\par  Example: Suppose ${\lambda_i}^a =  {\epsilon_i}^j (a) x_j$, such as a rotation generator (we may factor out the infinitesimal scale), for ${\epsilon_i}^j (a)$  antisymmetric constants. Then,   
$$[ {\lambda_i}^a {\partial \over \partial x_i}, {\lambda_j}^b {\partial \over \partial x_j}]
= x_j {M_i}^j(b,a) {\partial \over \partial x_i}, \eqno(8.14)$$
where
$$  {M_i}^j(b,a)= {\epsilon_i}^k (b){\epsilon_k}^j (a)-{\epsilon_i}^k (a){\epsilon_k}^j(b). \eqno(8.15)$$
For the rotation group, these form a finite Lie algebra.  The group
acts on the generating function (which forms a representation) but does not affect the $\{q,p\}$ variables. 
\bigskip
\noindent{\bf  9. Mapping of Bounded Submanifolds}
\bigskip
\par Since the mapping that we have constructed carries a Euclidean
phase space into a geometrical form, it is natural to study possibly
non-trivial topological properties that this geometrical space could
have. As a simple example, consider a potential in the Euclidean space
intwio dimensions which contains two identical finite depth potential
wells with lower bound $E_0$, and centers
spaced along the $x$-axis. Above a certain energy, say $E_1$, there is
just one connected region of motion, and between $E_1$ and $E_0$ there
are two separated regions. The total energy serves as a {\it height}
function, in the terminology of Morse theory [17]. 
\par Let us first consider a particle with energy $E_0 < E <E_1$. A
parricle in one of these wells has an orbit that is confined to this
well. If it reaches  the boundary where $E=V$, the momentum (and velocity)
vanishes, and the orbit necessarily then retraces its path as under time
reversal. Under the symplectomorphism,  this orbit is mapped into a
geodedsic curve,  and by the property of $1:1$ mapping, the corresponding geodesic curve must
stop and retrace its path as under time reversal as well. The family
iof all such orbits for a given value of $E$ defines a boundary in the
geometric space, and is therefore a closed submanifold with boundary. 
\par It is clear  that such orbits associated with each well (at a given value of $E$)
separately are disjoint since they are disjoint in the original
space. Increasing the energy above the value $E_1$ would result in a
single connected region for the geometric orbits.  Therefore the
homotopy classes of the possible orbits change as a function of the
height function $E$. We shall explore the consequences, in particular,
of the existence of topoligical invariants, in this context in a later
publication.
\bigskip
\noindent{\bf 11.Summary and Conclusions}
\bigskip   
\par  In this paper we have constructed a canonical transformation
from a Hamiltonian of the usual form $(1.1)$ to a geometric form $(1.2)$. 
\par We have given the basic mathematical
 formulation in terms of the geometry of symplectic manifolds.
\par For the central purpose of our construction, we
 formulate the process of studying stability by means of geodesic
 deviation in terms of geometric methods, making clear
 the relation between stability in the geometric manifold and the
 original Hamiltonian motion.
\par We then give an algorithm  for solving 
 the  nonlinear equations for the generating function of the canonical
 transformation.  This algorithm was then studied for the simple case of one
 dimension, and we proved convergence of the recursive scheme under certain reasonable assumptions.
 \par Since the series expansions generated by the algorithm for finding the solutions for the generating function may have a bounded domain of convergence, we studied  (in general dimension) the possiblity of
 shifting the origin in order to carry out the expansions based on a new origin. As for the analytic continuation of a function of a complex variable, this procedure can extend the solutions for the generating function to a maximal domain.  
\par Since the image space of the symplectomorphism has geometrical
 structure, it is natural to study its properties under local
 diffeomorphisms. A local change of variables $\{x,\pi\}\rightarrow \{x',\pi'\}$ (leaving the variables of the original space unchanged) alters the structure
 of the mapping from the original variables $\{q,p\}$ to the new variables $\{x',\pi'\}$;  we study the effect of infinitesinal diffeomorphims of this type on the generating function.
 \par We finally discussed briefly the mapping of bounded closed submanifolds,
 created by potential wells in
 the Hamiltonian space, corresponding to closed submanifolds in the
 geometric space, where Morse theory may be applied, to open the possibility of obtaining a new
 class of conserved quantities associated with homotopies of the
 image space.

\bigskip
\noindent References
\bigskip
\frenchspacing
\item{1.} Lawrence Horwitz, Yossi Ben Zion, Meir Lewkowicz, Marcelo Schiffer and Jacob Levitan,
Geomtry of Hamiltonian Chaos,  Phys. Rev. Lett. {\bf 98}, 234301 (2007).
\item{2.}Lapo Casetti and Marco Pettini, Phys. Rev. E {\bf 48}, 4320 (1993);  Marco Pettini, {\it Geometry and Topology in Hamiltonian Dynamics and Statistical Mechanics}, Springer, New York (2006), and references therein. See, in particular,  Carl Gustav Jacob Jacobi, {\it Vorlesungen \"uber Dynamik}, Verlag Reimer, Berlin 1884; Jacques Salomon Hadamard, J. Math. Pures Appl. {\bf 4}, 27 (1898). 
\item{3.} Vladimir I. Arnold, {\it Mathematical Methods of Classical Mechanics}, Springer-Verlag, New York (1978).
\item{4.}John Guckenheimer and Philip Holmes, {\it Nonlinear Oscillations, Dynamical Systems, and Bifurcations of Vector Fields}, Springer-Verlag, New York (1983). 
\item{5.} Emanuelle DiBenedetto, {\it Classical Mechanics, Theory and Mathematical Modelling}, Springer, New York (2011). See also, for example, Kathleen T. Alligood, Tim D. Sauer, and James A. Yorke, {\it Chaos, and Introduction to Dynamical Systems}, Springer, New York, (1996).
\item{6.} Martin C. Gutzwiller, {\it Chaos in Classical and Quantum,Mechanics}, Springer-Verlag, New York (1990).
 \item{7.} W.D. Curtiss and Forrest R. Miller, {\it Differentiable Manifolds and Theoretical Physics}, Academic Press, New York (1985).
\item{8.} Steven Weinberg, {\it Gravitation and Cosmology:
Principles and Applications of the General Theory of Relativity}, John
Wiley and Sons, New York (1972). 
\item{9.}Yossi Ben Zion and Lawrence Horwitz, Detecting order and chaos in three dimensional Hamiltonian systems by geometrical methods, Phys. Rev. E{\bf 76}, 046220 (2007), Yossi Ben Zion and Lawrence P. Horwitz, Phys. Rev. E {\bf 78}, 036209 (2008),
Yossi  Ben Zion and Lawrence P. Horwitz, Controlling effect of geometrically defined local structural changes on chaotic systems,  Phys. Rev. E {\bf 81} 046217 (2010).
\item{10.} Meir Lewkowicz, Jacob Levitan, Yossi Ben Zion, Lawrence P. Horwitz, Geometry of Local Instability in Hamiltonian Dynamics, in {\it Handbook of Chemical Physics (CMSIM)}, Chapman and Hall, CRC Press, Chap. 15, p. 231-252, May (2016).
\item{11.} Lawrence P. Horwitz, Asher Yahalom, Jacob Levitan and Meir Lewkowicz, An underlying geometrical manifold for Hamiltonian mechanics, Frontiers in Physics {\bf 12}, 124501 (2016).
\item{12.} Eran Calderon, Lawrence Horwitz, Raz Kupferman and Steven Shnider, On  the geometric formulation of Hamiltonian dynamics, Chaos {\bf 23} 013120 (2013).
\item{13.} See, for example, Ana Cannas da Silva, {\it Lectures on Sympletic Geometry}, Lecture Notes in Mathematics 1764, Springer, New York (2006).
\item{14.} Yossi Strauss, Lawrence P. Horwitz, Jacob Levitan and Asher Yahalom, Quantum field theory of Hamiltonian chaos, Jour. Math. Phys. {\bf 56} 072701 (2015).

\item{15.} Einar Hille, {\it Ordinary Differential Equations in the
Complex Plane}, John Wiley and Sons, New York (1976).
\item{16.}  T. Frankel, {\it The Geometry of Physics. An
Introduction} Cambridge University Press, Cambridge (1997). 
\item{17.} J. Milnor, {\it Morse Theory}, Annals of Mathematics
Studies 51,  Princeton University Press, Princeton (1969). See also [16].

\end